\documentclass[preprint, aps,prb,amsfonts,nobibnotes,superscriptaddress,showpacs]{revtex4-1}

\usepackage{dcolumn}
\usepackage{amsmath}
\usepackage{amssymb}
\usepackage{graphicx}
\usepackage{bm}
\usepackage{color}

\begin{document}

\title{{\normalsize A Versatile Setup for Ultrafast Broadband Optical Spectroscopy of Coherent Collective Modes in Strongly Correlated Quantum Systems}}

\vspace{5cm}

\author{Edoardo Baldini}
	\affiliation{Laboratory for Ultrafast Microscopy and Electron Scattering, IPHYS, \'Ecole Polytechnique F\'{e}d\'{e}rale de Lausanne, CH-1015 Lausanne, Switzerland}
	\affiliation{Laboratory of Ultrafast Spectroscopy, ISIC, \'Ecole Polytechnique F\'{e}d\'{e}rale de Lausanne, CH-1015 Lausanne, Switzerland}

\author{Andreas Mann}
	\affiliation{Laboratory for Ultrafast Microscopy and Electron Scattering, IPHYS, \'Ecole Polytechnique F\'{e}d\'{e}rale de Lausanne, CH-1015 Lausanne, Switzerland}

\author{Simone Borroni}
	\affiliation{Laboratory for Ultrafast Microscopy and Electron Scattering, IPHYS, \'Ecole Polytechnique F\'{e}d\'{e}rale de Lausanne, CH-1015 Lausanne, Switzerland}

\author{Christopher Arrell}
	\affiliation{Laboratory of Ultrafast Spectroscopy, ISIC, \'Ecole Polytechnique F\'{e}d\'{e}rale de Lausanne, CH-1015 Lausanne, Switzerland}
	
\author{Frank van Mourik}
	\affiliation{Laboratory of Ultrafast Spectroscopy, ISIC, \'Ecole Polytechnique F\'{e}d\'{e}rale de Lausanne, CH-1015 Lausanne, Switzerland}

\author{Fabrizio Carbone}
	\affiliation{Laboratory for Ultrafast Microscopy and Electron Scattering, IPHYS, \'Ecole Polytechnique F\'{e}d\'{e}rale de Lausanne, CH-1015 Lausanne, Switzerland}

\date{\today}

\begin{abstract}
A femtosecond pump-probe setup is described that is optimised for broadband transient reflectivity experiments on solid samples over a wide range of temperatures. By combining a temporal resolution of 45 fs and a broad detection range between 1.75 and 2.85 eV, this apparatus can provide insightful information on the interplay between coherent collective modes and high-energy electronic excitations, which is one of the distinctive characteristics of strongly interacting and correlated quantum systems.
The use of a single-shot readout CMOS array detector at frame rates up to 10 kHz allows to resolve coherent oscillations with amplitudes below 10$^{-4}$ in $\Delta$R/R. We demonstrate the operation of this setup on the prototypical charge-transfer insulator La$_2$CuO$_4$, revealing the presence of coherent optical phonons with frequencies as high as 13 THz.
\end{abstract}


\maketitle

\section{Introduction}

One of the most intriguing fields of research in contemporary condensed matter physics is the investigation of many-body effects in strongly correlated quantum systems. This class of materials provides an excellent playground for studying exotic phenomena involving charge, lattice, spin and orbital degrees of freedom and leading to extraordinarily varied chemical and physical properties. Understanding electronic correlations in prototypical systems like cuprates and manganites can open the doors to the potential design and engineering of novel materials with tailored functionalities. As a consequence, the main goal of current research is to identify the underlying mechanisms shared by these solids, so that these emerging phenomena can be accounted for by new theoretical models. Up to now, two distinctive features characterizing this class of solids have been identified by a number of spectroscopies: i) the presence of strong interactions (strong electron-boson coupling), which gives rise to phenomena like polaronic transport, incipient charge- and spin-density wave instabilities and renormalizations of the electronic structure~\cite{ref:shen}; ii) a non-trivial interplay between low- and high-energy scales, whose origin lies in the strong electronic correlations spreading the material's spectral-weight over a wide energy range~\cite{ref:basov}. Addressing such phenomenology is a challenge for experimental methods. Potentially, one needs an experimental technique that can disentangle the contributions of the different degrees of freedom and simultaneously monitor the low- (meV) and high-energy (eV) scales with high (tens of fs) temporal resolution.

In the last decades, ultrafast optical spectroscopies were established as an effective tool for revealing the temporal hierarchy of phenomena occurring in many-body systems upon photoexcitation. These techniques have been greatly improved since their early implementation on bulk semiconductors and nanostructures~\cite{ref:shank, ref:peyghambarian}, reaching a time resolution of tens of femtoseconds and capturing the ultrafast dynamics in previously unexplored spectral ranges~\cite{ref:kaindl, ref:wall, ref:pashkin, ref:baldini_TiO2}. Moreover, a unique strength of these methods is the possibility to access the dynamics and the intrinsic properties of specific low-energy Raman-active bosonic collective modes, which are coherently excited via the Impulsive Stimulated Raman Scattering (ISRS) process~\cite{ref:merlin} or by a long-lived perturbation of the electronic ground state~\cite{ref:zeiger}. 

\noindent The combination of the ISRS framework and the use of a broad window of detection in the optical range allows one to address a pivotal aspect of the paradigm behind strongly correlated quantum systems, \textit{i.e.} the coupling between Raman-active low-energy (tens of meV) bosonic modes and high-energy ($\sim$ eV) charge excitations on the Mott scale~\cite{ref:mansart_2012, ref:mann_2015, ref:rossi, ref:fausti, ref:borroni, ref:baldini_cuprate, ref:mann_2016}. When corroborated by theoretical calculations, this approach provides a very selective and quantitative estimate of the electron-boson coupling~\cite{ref:mann_2015, ref:rossi, ref:fausti}. In addition, when some external parameter (\textit{e.g.} temperature, pressure, magnetic field) is varied, the change of the electron-boson coupling can be followed through the phase diagram~\cite{ref:fausti}.

To assess this phenomenology, one needs to develop an instrument which is capable of: i) achieving a high time-resolution, to detect possibly excited low-energy coherent bosonic modes of the material under study; ii) offering a broad detection window covering the region of the high-energy interband transitions of the solid, where the bosonic modes are likely to resonate; iii) providing a high versatility in the determination of the nonequilibrium optical properties under varying experimental conditions; iv) achieving a high signal-to-noise ratio to clearly identify the spectro-temporal features characterizing the ultrafast optical response. Although a similar experimental approach has been largely used in the past in the fields of molecular spectroscopy and semiconductor physics~\cite{ref:polli, ref:megerle, ref:aubock_1, ref:aubock_2}, the extension of this framework to the study of complex solids requires the development of a suitable cryogenic environment for the sample and, in most situations, the implementation of a reflection geometry.

\noindent Taking inspiration from pioneering room-temperature works on bulk noble metals~\cite{ref:schoenlein, ref:sun, ref:delfatti, ref:kruglyak}, first important steps towards this goal have been made in the last years. A 250 kHz repetition-rate setup has been used for measuring the broadband transient-reflectivity ($\Delta$R/R) of cuprates~\cite{ref:fausti, ref:giannetti_2011, ref:dalconte_2012, ref:cilento, ref:novelli_2014} and other transition metal oxides~\cite{ref:novelli_2012, ref:randi}, offering a high tunability in terms of pump photon energies and explored temperature range. However, due to limitations in the time-resolution, coherent bosonic modes were only observed up to a frequency of $\sim$4.5 THz~\cite{ref:fausti}. More recently, a broadband (0.75 - 2.40 eV) and high time resolution (9 - 13 fs) instrument has been implemented with the same purpose~\cite{ref:dalconte_2015}. Nevertheless, up to now, only operation down to 100 K has been demonstrated, due to the challenge of maintaining the same pulse compression inside a cryostat. This hinders the possibility to probe the strongly correlated quantum phases emerging at a lower temperature scale. 

In this work, we present a versatile experimental setup for femtosecond broadband optical spectroscopy in the visible spectral range, which allows us to specifically address this problem. The apparatus is based on a cryo-cooled amplified Ti:Sapphire laser and offers an overall time resolution of $\sim$ 45 fs. The pump beam can be tuned to different photon energies to explore several excitation conditions, while the probe is a broadband continuum covering the spectral region from 1.75 to 2.85 eV. The setup has been designed to allow systematic transient reflectivity ($\Delta$R/R) and transmissivity ($\Delta$T/T) studies for different temperatures (8 - 340 K) and applied magnetic fields (0 - 1 T). Section II of this article describes the laser system and the optical design of the setup. Section III illustrates the design of the cryostat in which the experiments are performed. Section IV highlights the operation of our instrument on a single-crystal of La$_2$CuO$_4$ as a benchmark solid system. We observe the rise of an ultrafast (resolution-limited) $\Delta$R/R response and the emergence of coherent oscillations in the time domain, which are ascribed to coherent optical phonons with a frequency as high as $\sim$ 12.9 THz. The basic scheme of this setup opens up intriguing possibilities for the future implementation of low-temperature time-resolved broadband magneto-optical measurements and time-resolved spectroscopic ellipsometry~\cite{ref:boschini}.
\newpage

\section{Experimental setup}

\subsection{Laser system and electronics}

Our femtosecond broadband optical spectroscopy setup is laid out in a standard pump-probe configuration. An overview of the setup and its laser system is given in Fig. 1. A Ti:Sapphire oscillator (KM Labs, Griffin), pumped by a continuous-wave Nd:YVO$_4$ laser (Coherent, Verdi-V5), emits $\sim$ 45 fs pulses at 1.55 eV (800 nm) with a repetition rate of 80 MHz. The output of the oscillator seeds a cryo-cooled Ti:Sapphire amplifier (KM Labs, Wyvern-1000), which is pumped by a Q-switched Nd:YAG laser (Lee Laser, LDP-200MQG). This laser system provides $\sim$ 45 fs pulses at 1.55 eV (800 nm) with a repetition rate of 3 - 10 kHz at a pulse energy of up to 3 mJ~\cite{ref:ojeda}.

\noindent In a basic configuration, a few $\mu$J of the laser output are used for the pump beam when exciting a sample at 1.55 eV. Optionally, the pump output can be frequency-doubled to 3.10 eV (400 nm) using a $\beta$-barium borate (BBO) crystal, or can be tuned in the infrared between $\sim$ 60 meV (20 $\mu$m) and 1 eV (1.2 $\mu$m) using a commercial optical parametric amplifier (Light Conversion, TOPAS-C with NDFG stage). The white light pulse serving as the probe beam is generated by focusing pulses with an energy of about 1 $\mu$J into a calcium fluoride (CaF$_2$) crystal of 3 mm thickness using a combination of a lens with short focal distance and an iris to limit the numerical aperture of the incoming beam. The generated continuum ranges from 1.72 to 2.92 eV and is peaked around 2.20 eV. The CaF$_2$ crystal is mechanically oscillated to slow down crystal degradation by spatial hole-burning caused by the high laser intensity. The residual component of the generating beam at 1.55 eV is eliminated using a high-pass colored glass filter, and the divergent white light beam is collimated and focused onto the sample in a dispersion-free manner using a pair of off-axis parabolic mirrors, impinging onto the sample surface at an angle of 15-20$^{\circ}$.

\noindent The spot sizes of the pump and probe beams at the sample surface are measured by a CCD-based beam profiler. Typical full width at half maximum (FWHM) dimensions of the near-Gaussian profiles are 150 x 150 $\mu$m$^2$ for the pump beam and 50 x 50 $\mu$m$^2$ for the probe beam. The polarizations of the pump and probe beams can be adjusted using half-wave retardation plates.

\noindent The time-delay between pump and probe is adjusted via the use of a retroreflector mounted on a mechanical delay stage (Newport, M-UTM25PE-1), which is installed in the probe path prior reaching the CaF$_2$ crystal for white light generation.

\noindent The specular reflection of the probe beam is collimated into an optical fiber using an achromatic lens. The fiber couples the beam into a \textit{f}/4-spectrometer, which uses a linear complementary metal oxide semiconductor (CMOS) array as a detector (Hamamatsu, S10453-1024Q). The electronic shutter of the CMOS detector is synchronized to the incoming laser pulses through a series of electronic circuits, sketched in Fig. 2. The core piece is a 16-bit analog-to-digital converter (ADC) that produces a master clock pulse train at 11 MHz, fast enough to read out all 1024 detector pixels from pulse-to-pulse at the maximum laser repetition rate of 10 kHz. The signal derived from the amplifier pump laser trigger is synchronized to the master clock and ultimately paces the readout operations of the CMOS array. The same trigger signal is halved in frequency and sent to the chopper controller, which synchronizes the rotation of the mechanical chopper blade to the incoming laser pulses. The chopper runs at a quarter of the laser repetition rate, blocking and letting pass pulses in pairs of pump pulses. This is done to eliminate a particular intensity fluctuation that often occurs with continuously pumped Q-switched pump lasers that are operated at repetition rates around the inverse lifetime of the gain medium (in Nd:YAG, $\mathrm{T_f}$ = 230 $\mu$s). In this regime, the ``memory" of the gain medium extends over multiple pulses. Due to feedback between pulses (a relatively weak pulse leaves more residual gain for the next  pulse, which will be stronger and leave less residual gain for its successor) repetitive intensity fluctuations are observed at half the repetition rate, often referred to as a ``Ding-Dong" effect. Even when this modulation is very minor ($\ll$ 1\%), it can give rise to a strong artefact in the measured transient signal. By chopping pulse pairs (i.e. chopping at 1/4 of the repetition rate) this artefact is fully suppressed.

\noindent A gate signal is then created using the reference output of the chopper controller, and combined with the trigger signal. This gated trigger is finally used to start the data acquisition process of the CMOS array. Along the way, several extra delays are added to compensate for cable lengths, beam propagation, and positioning of the various elements along the beam path. Using this process, it is ensured that: i) the electronic shutter of the CMOS array is opened in a time window around the arrival of the laser pulses, ii) the shutter is closed between the arrival of pulses to reduce background noise, and iii) all detector pixels are read out in the period between pulses. The synchronization of the detector is further explained elsewhere \cite{ref:hamamatsu}. To compensate for the pump beam fluctuations, a photodiode is also connected to a channel of the ADC to monitor and eliminate the fluctuations of the pump beam intensity.
\newpage

\section{Sample environment}

The cryostat assembly used for optical experiments, shown in Fig. 3, is based on a closed-cycle liquid helium cold head (Advanced Research Systems Inc., DE-204) with a vibration-reducing gas interface. The cryostat expander is mounted to the laboratory ceiling, while the cold head and cryostat shroud are supported from the optical table, completely isolating the latter from the strong vibrations of the expander. The cold head allows for experiment temperatures between 8 K and 340 K. Samples are mounted using fast-drying silver paint on a small copper plate attached to a copper wire of 2 mm thickness descended from the cold head. 

The sample shroud is a custom design made of aluminium, allowing for the application of an external magnetic field, and is fitted to a standard ConFlat flange using a soft, annealed copper gasket. Optical access is provided via a 1-inch window port at the front. The window material is chosen according to the pump and probe energies of the experiment. For experiments in the visible/near-infrared range, CaF$_2$ is used. A turbo pump, backed up by an oil pump, is attached to the back of the shroud, the proximity to the sample significantly improving the vacuum. The vacuum pressure is measured by a sensor attached to the cold head shroud. Achievable minimum pressures range from 10$^{-8}$ mbar at room temperature to 10$^{-9}$ mbar at 10 K. The vacuum is improved after closing the cryostat by heating up to the maximum temperature over $\sim$ 10 hours.

The sample shroud depicted in Fig. 3 is optimized for transient-reflectivity and ultrafast magneto-optical Kerr effect measurements. The small cubic sample space of about 1 cm edge length allows bringing the poles of an electromagnet (GMW Associates, 3470) close enough to the sample to achieve in-plane fields ranging from 0 to 1 T. For transmission and ellipsometric measurements, different aluminium shrouds were designed. In particular, for transient spectroscopic ellipsometry, the shroud features probe entrance and exit windows under 70$^{\circ}$ with respect to the pump entrance window (which allows the pump to illuminate the sample under normal incidence).
\newpage

\section{Measurements}
\subsection{Signal acquisition and processing}

In Section III we described the possibility of performing measurements both in reflection ($\Delta$R/R) and transmission ($\Delta$T/T) geometry, depending on the specific sample (bulk crystal, thin-film...) that is measured. In this Section, we focus on the acquisition and preliminary processing of the data. For simplicity, we restrict our discussion to transient-reflectivity measurements on bulk crystals.

\noindent The differential reflectivity change $\Delta$R/R is extracted from the data by calculating the quantity $\Delta$R ($\omega$, t):

\begin{equation}
\frac{\Delta R}{R}(\omega, t) = \frac{R_{pumped}(\omega, t) - R_{unpumped}(\omega, t)}{R_{unpumped}(\omega, t)}
\end{equation}
\vspace{0.1cm}

\noindent where $R_{pumped}$ and $R_{unpumped}$ are the two arrays containing the sums over the spectra received from the ADC. Every individual curve is corrected for the offset spectrum $R_{dark}$($\omega$), which is acquired before every scan by measuring the light entering the fiber when the probe beam is blocked and the pump beam is open, eliminating the scattered pump light as well as any ambient light from the acquired spectra. It is important to note that the sign of the right-hand side of Eq. (1) depends on the electronic phase offset of the chopper as well as its position along the beam path, which have to be adjusted properly to obtain the physical sign of the reflectivity change. Finally, the dataset $\Delta$R/R($\omega$,t) containing the reflectivity change at every time delay for every probe energy is retrieved in the form of a rectangular matrix. In a typical experiment, the acquisition of each matrix is repeated multiple times to improve the statistics of the measurement. Hence, the experiment strongly relies on the repeatability of scans, impliying the stability of the sample under laser light illumination for hours. 

Acquiring a dataset for one set of experimental parameters (fluence, polarization, sample temperature) typically requires from few hours to a whole day, depending on the signal level and the desired signal-to-noise ratio. Before the data are analyzed, the $\Delta$R/R matrix has to be corrected for the Group Velocity Dispersion (GVD) of the probe beam. Since the probe beam is not dispersion-compensated after generation of the white light continuum, the probe pulses arrive at the sample stretched to a duration of few ps. This is beneficial for the experiment, because it significantly reduces the momentary probe light intensity in the sample. It is noteworthy that the probe beam dispersion is not a limiting factor for the time-resolution of the setup, which is given on the detection side by the much smaller effective pulse duration per detector pixel. The correction of the GVD is done numerically by defining values for time-zero (\textit{i.e.} pump-probe overlap) for a number of probe photon energies across the spectrum by looking at the time traces $\Delta$R/R(t). Typically, the onset of the signal can be used as point of overlap. The final result depends on the exact method used to define time-zero given a certain slope, but consistent results can easily be achieved due to the fact that the GVD can be represented as a Taylor series with sizable contributions only up to third order. While the time traces can be understood in terms of the sample response to a $\delta$-like excitation convolved with the Gaussian shape of the pump pulse, it is in general not necessary to deconvolve the traces to obtain the true pump-probe overlap in time. It should be noted, however, that the GVD correction has an intrinsic uncertainty that can not be pushed very far below half the pump pulse duration. The raw data are corrected for GVD by shifting each time trace by its assigned time-zero value, as well as subtracting any possible offsets caused by noise in the offset spectrum from Eq. (1). The matrix is then trimmed around the edges to eliminate missing data points.

Integrating over a typical value of 1000 laser shots per acquisition, the setup has an intrinsic noise level of about 0.1$\%$ RMS. The main noise sources are given by the amplifier output noise, which is strongly increased by the nonlinear white light generation process, the electronic noise of the CMOS array and the noise due to AD conversion. The signal-to-noise ratio is improved statistically by repeating each measurement many times, typically up to 100 scans per matrix. In addition, the output of several detector pixels is binned, averaging in energy. In this way, relative reflectivity changes down to the order of 10$^{-4}$ can be observed. Although the measurable signal variation is several orders of magnitude higher than the one detected in single-wavelength (i.e. oscillator based) pump-probe experiments, we remark that the advantage of our broadband probe is to provide a spectrally-resolved picture of the nonequilibrium optical response, revealing valuable information on the electronic structure of the material under study. This aspect is particularly crucial when one aims at unveiling spectral weight transfers occurring on ultrafast timescale in strongly correlated quantum systems. In addition, for most systems, the pump fluences required for producing a reflectivity change of 10$^{-4}$ in the optical range are still within the linear regime of the sample response~\cite{ref:giannetti_2009}. Finally, the use of an amplified laser system is ideal for the study of photoinduced phase transitions in complex materials, where fluences on the order of $\sim$ mJ/cm$^2$ are typically required~\cite{ref:cavalleri, ref:gedik, ref:hilton, ref:beaud, ref:eichberger}.

\subsection{Application on La$_2$CuO$_4$}

To demonstrate the capabilities of our setup and its intrinsic time-resolution, we perform broadband transient-reflectivity on a single-crystal of La$_2$CuO$_4$. This material is the undoped parent compound of the cuprate series La$_{2-x}$(Sr,Ba)$_x$CuO$_4$ and it represents a prototypical example of a strongly correlated charge-transfer insulator. As a consequence, below the N\'eel temperature $\mathrm{T_N}$ (which is 300 K in our sample), the magnetic moments of Cu atoms order antiferromagnetically on the CuO$_2$ planes.

\noindent In our measurement, we pump a (010)-oriented single-crystal of La$_2$CuO$_4$ with 45 fs pulses centered around 3.10 eV. The probe is a broadband continuum extending from 1.72 eV to 2.92 eV. The pump polarization is parallel to the [100] crystallographic direction, while the probe polarization is set along the [001] direction. Hence, the pump pulse excites particle-hole pairs across the charge-transfer gap of the in-plane response (i.e. on the CuO$_2$ planes), while the probe monitors the optical response of the material along the c-axis. Given our pump pulse duration, Raman-active modes with a frequency up to $\sim$ 22 THz can be coherently excited~\cite{ref:merlin, ref:zeiger}. The laser system repetition rate is 3 kHz and the pump fluence absorbed by the sample is $\sim$ 4.4 mJ/cm$^2$. The detailed analysis of the spectro-temporal evolution of the $\Delta$R/R signal and its relevance for the physics of La$_2$CuO$_4$ will be the subject of a separate paper. Here, we only report on the observation of an ultrafast resolution-limited rise in the $\Delta$R/R response and on the evidence for high-energy coherent optical phonons present over the wide probe range. Fig. 4(a) displays the raw color-coded map of $\Delta$R/R at 10 K as a function of probe photon energy and time-delay between pump and probe, as obtained after the measurement. The map is then corrected for the GVD and the background noise prior analyzing the data. In Fig. 4(b) the rise of the temporal trace at 2.60 eV (integrated over a spectral region of 0.20 eV) is shown. We observe an ultrafast rise of $\sim$ 50 fs in the transient response, which is a clear fingerprint of the time-resolution of our experimental setup. In Fig. 4(c) we demonstrate that the combination of our high time-resolution and of the correct polarizations of pump and probe beams is crucial for revealing the emergence of coherent collective modes in the measured response. Indeed, by selecting a temporal trace around the probe photon energy of 2.00 eV and integrating it over a spectral range of 0.40 eV, we can clearly identify a complex oscillatory signal. This is best highlighted in the inset of Fig. 4(c), where the residuals from a multiexponential fit of the temporal trace at 2.00 eV are reported. Given the absence of a clear periodicity of the observed modulations, we can associate the coherent response to the beating among several coherent bosonic modes. Fig. 4(d) shows the Fourier transform of the residuals reported in the inset of Fig. 4(c). From the Fourier transform we can observe the presence of five peaks at 3.68 THz (I), 4.55 THz (II), 6.93 THz (III), 8.23 THz (IV) and 13 THz (V). The uncertainty on the frequency of such modes is $\sim$ 0.22 THz and results from the measured time window in the pump-probe data ($\sim$ 4.6 ps). Remarkably, all these modes have a clear counterpart in spontaneous Raman scattering measurements of the same compound, being totally-symmetric ($\mathrm{A_g}$) phonon modes~\cite{ref:nimori}. The emergence of these modes in different regions of our probed spectrum is the fingerprint of their different coupling with the high-energy excitations of the solid. The fine details of the electron-phonon coupling will be estimated by extracting the Raman matrix elements (\textit{i.e.} the spectral evolution of the amplitudes) of these modes and by their direct comparison with the theoretical Raman matrix elements computed via \textit{ab initio} calculations.
\newpage

\section{Conclusion}

In this work, we have presented the performances of an experimental setup for ultrafast broadband optical spectroscopy covering the visible spectral range. At the lowest level, this technique can be used to develop realistic nonequilibrium models for the dielectric function of strongly correlated quantum systems. At the highest level, the combined time-resolution (45 fs) and wide detection window (1.72 - 2.92 eV) allow the study of coherent collective bosonic modes and of their impact on the high-energy scale of these materials. We showed the reliability of our setup by measuring a single-crystal of La$_2$CuO$_4$ and revealing the presence of coherent optical phonons with a frequency as high as 13 THz. This first demonstration paves the route for a future systematic study of coherent phenomena in complex solids and represents a fundamental step prior to the application of more sophisticated structural probes like ultrafast x-ray and electron diffraction.

\begin{acknowledgments}
E.B. and A.M. contributed equally to this work. We thank K. Conder and E. Pomjakushina for providing the La$_2$CuO$_4$ sample and J. Lorenzana for the useful discussions. Work at LUMES and LSU was supported by NCCR MUST and by the ERC starting grant USED258697.
\end{acknowledgments}
\newpage

\bibliography{Papers}
\newpage

\begin{figure}[h]
\begin{center}
\includegraphics[width=0.9\columnwidth]{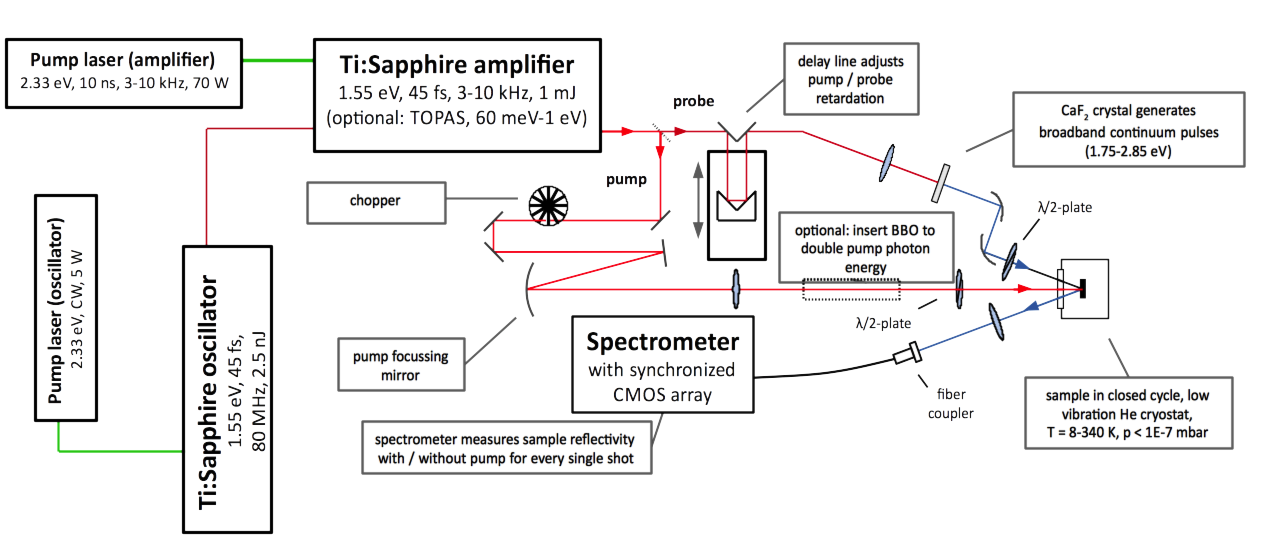}
\caption{A diagram of the experimental setup, detailing the laser system and the broadband pump-probe experiment.}
\label{fig:Fig1}
\end{center}
\end{figure}

\begin{figure}[h]
\centering
\includegraphics[width=0.8\columnwidth]{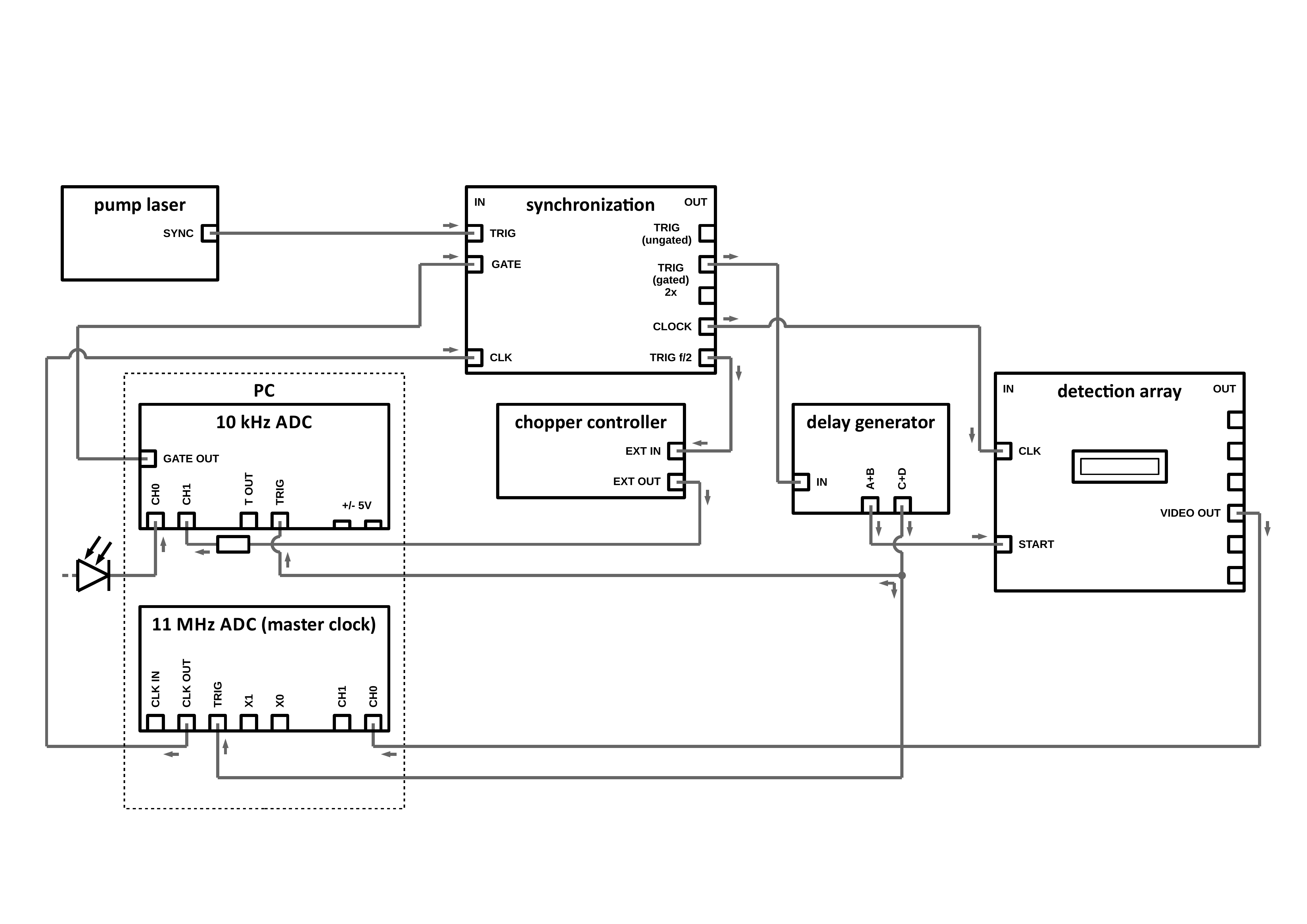}
\caption{Schematic representation of the synchronization electronics.}
\label{fig:Fig2}
\end{figure}

\begin{figure}[h]
\centering
\includegraphics[width=\columnwidth]{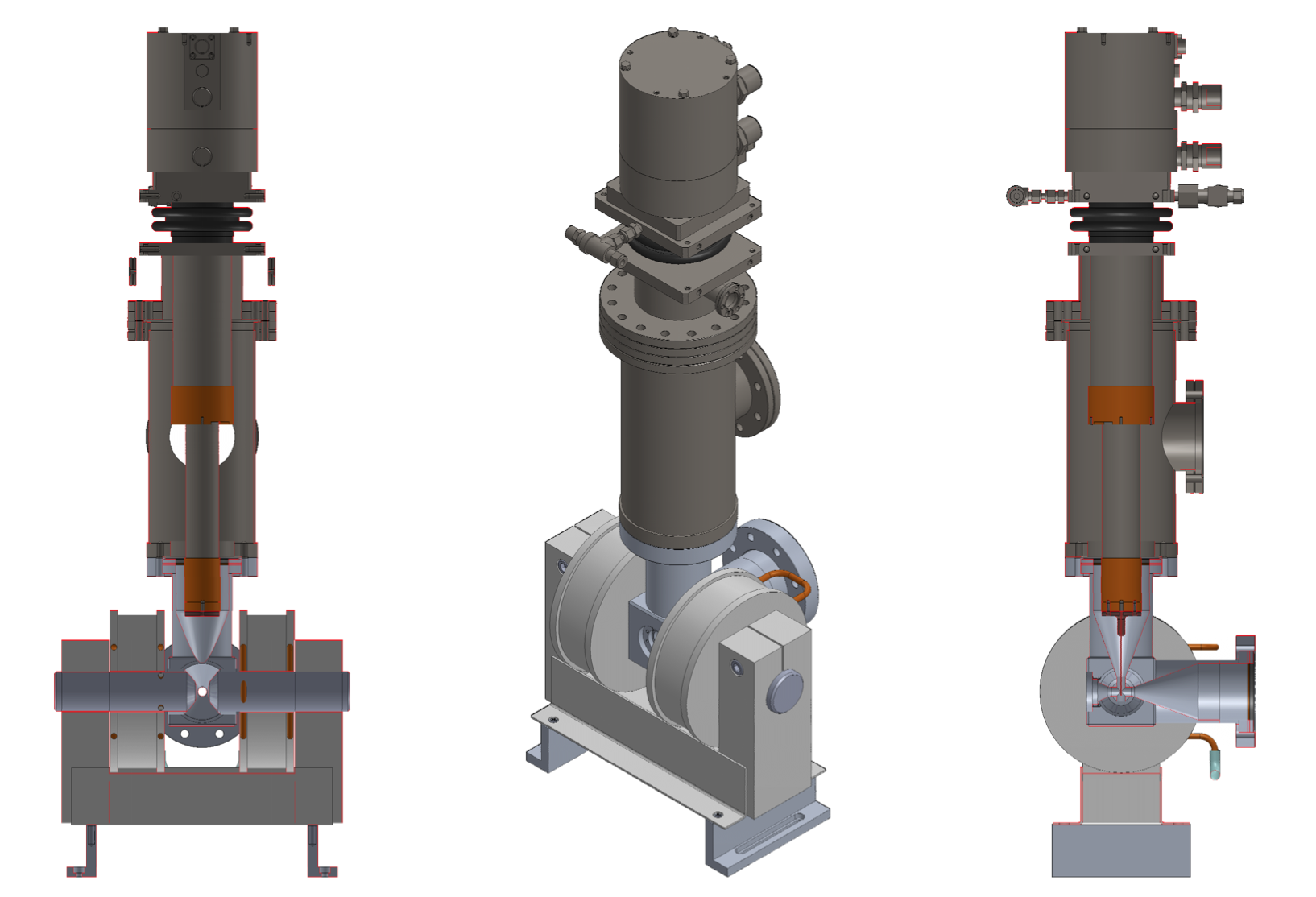}
\caption{The cryostat assembly with the magnet in position. From top to bottom, the assembly comprises the expander (dark gray), the gas interface (black), the cold head shroud (dark gray), and the aluminium sample shroud (light gray).}
\label{fig:Fig3}
\end{figure}

\begin{figure}[h]
\centering
\includegraphics[width=0.8\columnwidth]{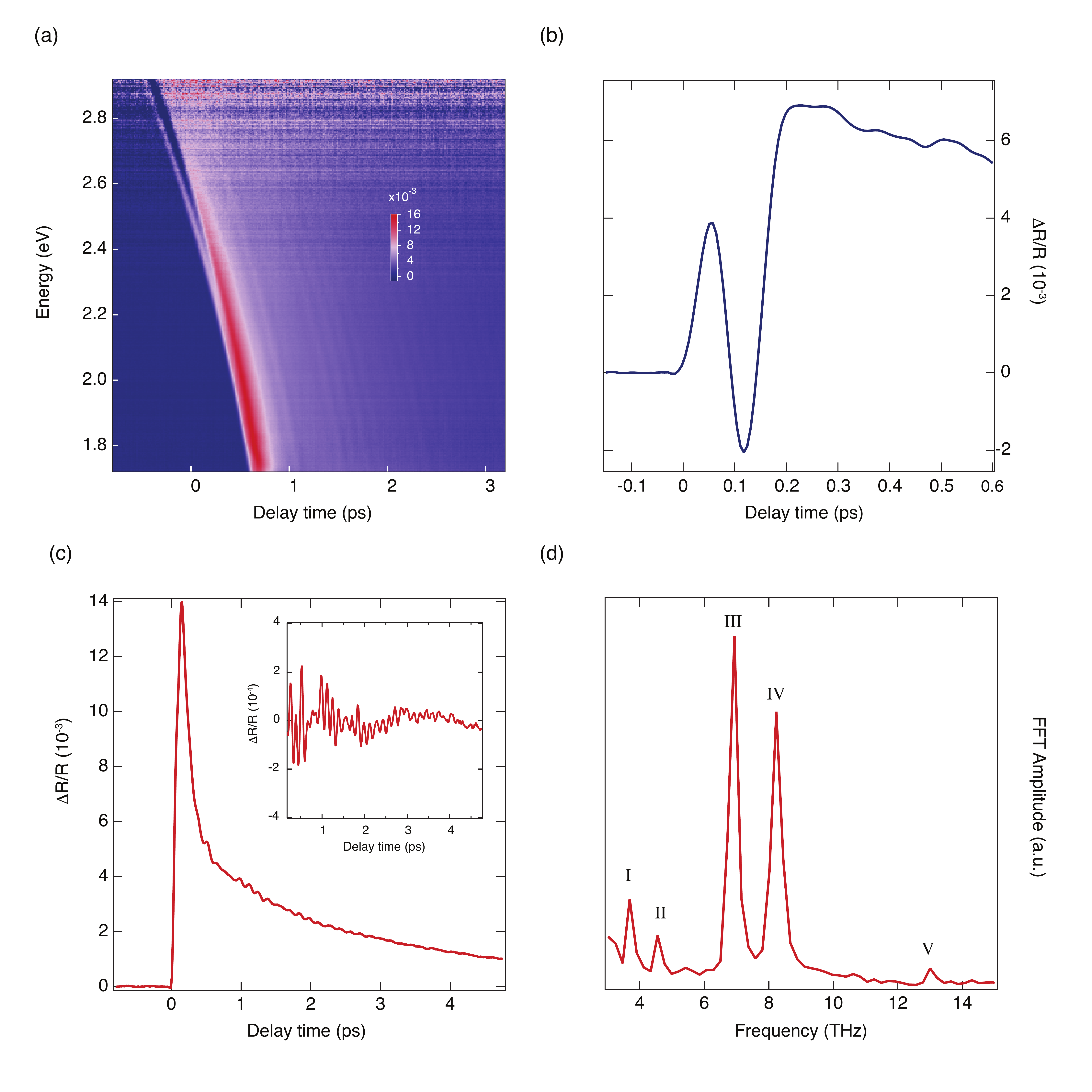}
\caption{(a) Raw color-coded map of the transient-reflectivity ($\Delta$R/R) of La$_2$CuO$_4$ at 10 K as a function of probe photon energy and time delay between pump and probe. The pump fluence absorbed by the sample is $\sim$ 4.4 mJ/cm$^2$. The pump and probe polarizations are parallel to the [100] and [001] crystallographic directions, respectively. (b) Rise of the temporal trace at 2.60 eV, integrated over 0.20 eV. (c) Temporal trace of $\Delta$R/R at a probe photon energy of 2.00 eV, integrated over 0.40 eV. Inset: residuals from a multiexponential fit of temporal trace at 2.00 eV. (d) Fourier transform analysis of the residuals in the inset of panel (c). The symbols I, II, III, IV, V identify peaks corresponding to the $\mathrm{A_g}$ phonon modes of the crystal.}
\label{fig:Fig4}
\end{figure}
\end{document}